\documentclass[pdflatex,sn-mathphys-num]{sn-jnl}

\usepackage{graphicx}%
\usepackage{multirow}%
\usepackage{amsmath,amssymb,amsfonts}%
\usepackage{amsthm}%
\usepackage{mathrsfs}%
\usepackage[title]{appendix}%
\usepackage{xcolor}%
\usepackage{textcomp}%
\usepackage{manyfoot}%
\usepackage{booktabs}%
\usepackage{algorithm}%
\usepackage{algorithmicx}%
\usepackage{algpseudocode}%
\usepackage{listings}%
\usepackage{bibunits}
\usepackage{comment}
\defaultbibliographystyle{unsrt}

\theoremstyle{thmstyleone}%
\theoremstyle{thmstyletwo}%

\theoremstyle{thmstylethree}%

\begin{document}

\title[Article Title]{Supercurrent spin Hall effect enabled nanopillar Josephson diodes}


\author[1,2]{\fnm{Debashree} \sur{Nayak}}\email{debashree.nayak@niser.ac.in}

\author[1,2]{\fnm{Dimple} \sur{Rani}}\email{dimple.rani@niser.ac.in}

\author[1,2]{\fnm{Prasanjit} \sur{Samal}}\email{psamal@niser.ac.in}

\author[1,2,3]{\fnm{Kartik} \sur{Senapati}}\email{kartik@niser.ac.in}

\affil[1]{\orgdiv{School of Physical Sciences}, \orgname{National Institute of Science Education and Research (NISER) Bhubaneswar}, \orgaddress{ \city{Jatni}, \postcode{752050}, \state{Odisha}, \country{India}}}

\affil[2]{\orgdiv{Homi Bhabha National Institute}, \orgname{Training School Complex}, \orgaddress{\street{Anushakti Nagar}, \city{Mumbai}, \postcode{400094}, \country{India}}}

\affil[3]{\orgdiv{Center for Interdisciplinary Sciences}, \orgname{National Institute of Science Education and Research (NISER) Bhubaneswar}, \orgaddress{\city{Jatni}, \postcode{752050}, \state{Odisha}, \country{India}}}

\begin{bibunit}[unsrt]

\abstract{ In the recent years it has been possible to achieve diode-like, non-reciprocal current-voltage response in Josephson junctions, despite the intrinsic symmetry of the Josephson effect itself. This is typically achieved by incorporating Rashba spin-orbit coupling into the Josephson junction as a strong inversion symmetry breaking component, and external magnetic field as a tuneable time-reversal symmetry breaking component. However, the efficiencies of the external field tuneable Josephson-diodes have remained limited to less than 10 \%, often measured below 100 mK temperature. In this work we take a new approach where non-reciprocity is induced by intrinsic SOC in a heavy metal Josephson barrier via the predicted supercurrent spin-Hall effect. By measuring a series of Nb-Pt-Nb nanopillar junctions we demonstrated field tuneable Josephson diode efficiencies as high as 17\%, measured above liquid Helium temperature. This was possible by the realization of a net non-equilibrium spin segregation in the Pt barrier, due to the supercurrent spin-Hall effect in the Pt barrier, analogous to the normal spin-Hall effect. As the direction of the induced spin moment is determined by the bias current, an external magnetic field causes the associated phases to add with opposite signs for opposite current directions, resulting in a nonreciprocal supercurrent across the junction.}




\maketitle
\section*{Main}

Josephson junctions are the quintessential non-linear devices which drives the entire field of superconducting quantum electronics. Therefore, a non-reciprocal version of this superconducting device, analogous to the semiconductor diode, is highly desirable circuit element for superconducting electronics. Asymmetric current–voltage characteristics in superconducting tracks and Josephson junctions is generally achieved by the explicit breaking of the time-reversal symmetry and the inversion symmetry for the Cooper pairs. Such conditions are intrinsically realized in non-centrosymmetric superconductors with strong spin–orbit coupling (SOC) \cite{Tokura2018,Nadeem_natRev,noncentrosymmetric_superconductor_SDE_exp,noncentrosymmetric_superconductor_SDE_Th}. However, simultaneous breaking of inversion and time reversal symmetries can also be achieved in engineered heterostructures combining superconductors, ferromagnets, and materials with strong SOC \cite{L_Fu_SDE,S/F_SDE_th,Ono_SF_multilayer,ando_SDE_Nb/V/Ta, RSOCdiodeAlAs-InAsplanar,RSOCdiodeBergerate,Rsocfinitemomentum1, unisersalJDE}. Various implementations of the above principle have been reported in the past few years to realize non-reciprocal Josephson junctions or Josephson diodes, including magnetic field-free \cite{Nb3Br8_Vanderwaal_JDE_Zerofiled, FieldfreeJDE1,jeon_Pt_YIG_0FJDE,fieldfreeJDE,fieldfreeJDE_Avradeep_pal} and field controlled\cite{NiTe2_topology_JJ, Strunk_2ndharmonic_JDE, Rsocfinitemomentum1,costa0-pidiode,Al_InAs_JDE_Javad_Shabani} variants. The field-free JJ diodes have relied on specialized barrier materials such as\cite{Nb3Br8_Vanderwaal_JDE_Zerofiled} Nb$_3$Br$_8$ and various topological quantum materials\cite{trilayer_grapgene_zerofield_JDE,Topology_JDE,Topology_JDE2,topology_JDE3,topology_JDE4}. NbSe$_2$/Nb$_3$Br$_8$/NbSe$_2$ Josephson junctions, where the barrier is an Obstructed Atomic Insulator (OAI), showed Josephson diode effect (JDE) due to asymmetric tunneling of the critical current $I_c$\cite{Nb3Br8_Vanderwaal_JDE_Zerofiled}. JJs fabricated with non-magnetic topological barriers have also resulted in JDE due to non-trivial band topology \cite{trilayer_grapgene_zerofield_JDE,Topology_JDE,Topology_JDE2,topology_JDE3,topology_JDE4}. In a recent experiment, Jeon et al.\cite{jeon_Pt_YIG_0FJDE} have shown Josephson diode effect in Nb-(Pt/YIG)-Nb planar junctions, where YIG proximity induced magnetism in the Pt layer resulted in finite-momentum Cooper pairing leading to asymmetric current-voltage curves in the absence of external magnetic field. Apart from these field-free devices, magnetic field controlled Josephson diodes have attracted considerable attention due to the possibility of in-situ tunability in quantum circuits. In these devices, generally a magnetic field is used as the time-reversal symmetry breaking agent, while the inversion symmetry-breaking is achieved by strong interfacial Rashba spin-orbit coupling \cite{Strunk_2ndharmonic_JDE, Rsocfinitemomentum1,costa0-pidiode,Al_InAs_JDE_Javad_Shabani,Turini2022}.  The interfacial Rashba SOC and the bulk Dresselhaus SOC results in spin momentum locking\cite{spin_mom_locking,Manchon2015} which naturally brings in non-reciprocity into the Josephson devices \cite{Amundsen_rev}. However, the performance of these Josephson diodes, in terms of the diode efficiency and the operating temperature, has remained very limited. A selected list of typical Josephson diode efficiencies and the respective operating temperatures is given in the supplementary Table-I. As evident from this table, maximum Josephson diode efficiencies of the order of 20$\%$ has been achieved at temperatures below 30 mK. A pertinent question is, therefore, whether the widely used Rashba SOC is practically limiting the achievable diode efficiency in JJs by imposing a planar geometry. In other words, can magnetic field controlled non-reciprocity be engineered into Josephson junctions without invoking the Rashba or Dresselhause type SOC, with better diode efficiencies at higher temperatures ?

In this work we demonstrate that intrinsic spin-orbit coupling, in heavy metal barriers, can naturally integrate magnetic field tunable supercurrent non-reciprocity, into fully metallic Nb-Pt-Nb vertical Josephson junctions. The emergence of Josephson diode effect (JDE), in this case, can be directly linked to a supercurrent generated effective spin moment \cite{spinhallinJJ, Hikino_tripletJJ} in the spin-orbit coupled barrier, analogous to the spin-Hall effect\cite{SHE_Pt3,SHE_Pt2,SHE_Pt,SHE_Pt1}. In the presence of an external in-plane magnetic field, the supercurrent spin-Hall effect (SSHE) induced spin magnetization can either add or subtract a relative phase to the junction upon the reversal of the bias current, thereby resulting in a critical current asymmetry. By measuring a series of Nb-Pt-Nb vertical Josephson nano-junctions we find strong experimental evidence for the presence of SSHE induced spin moment in the Pt layer, which is pivotal for the realization of JDE in the Nb-Pt-Nb Josephson junctions.    

\section*{Device geometry and principle of Josephson diode operation}
\begin{figure}[b]
    \centering
    \includegraphics[width=1\textwidth]{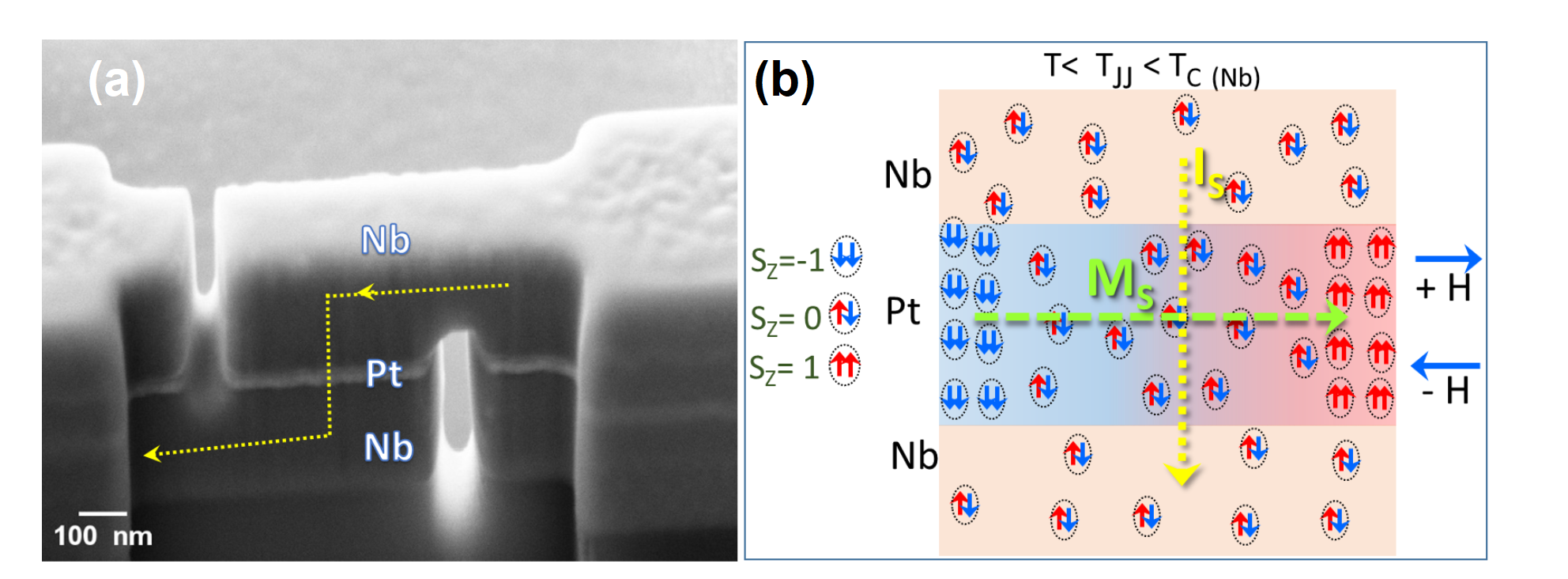} 
    \caption{\textbf{Origin of JDE from supercurrent Spin Hall effect}(a) Scanning electron micrograph of the junction architecture shown for a Nb(160\,nm)/Pt(30\,nm)/Nb(160\,nm) device. The dotted arrow indicates the current path across the nano-pillar junction (b) Schematic illustration of the accumulation of the S$_z$=-1 and S$_z$=+1 projections of the long range triplet component of the supercurrent, arising out of the interaction of the spin singlet Cooper pairs with strong SOC in Pt. This effect leads to a spin moment M$_S$ dictated by the magnitude and direction of the bias current I$_S$. The phase difference between the superconducting Nb electrodes is determined by the net effect of the external field H and the spin moment M$_S$. Therefore, a field-dependent non-reciprocity in the Josephson current can naturally emerges in these junctions. }
    \label{Fig.1}
\end{figure}
The design and the concept of the SSHE driven Josephson diode are described in the Figure 1. The Junction is a fully metallic vertical trilayer nano-pillar structure in which Nb superconducting electrodes are coupled via metallic Pt barrier, as shown in the scanning electron micrograph of a device in the Fig 1(a). In this geometry the Rashba contribution can be neglected as the interfacial electric field $\hat{e}$ (at Nb-Pt and Pt-Nb interfaces) is collinear with the direction of current injection into the junctions. In addition, the centrosymmetric crystal structure of Pt barrier rules out any contribution of the Dresselhaus SOC in the bulk of the junctions\cite{Dresselhaus,DSOC_noncentrosymmetric}. However, the presence of high atomic spin-orbit coupling in Pt barrier can lead to the Josephson diode effect by facilitating two different phenomena. Firstly, in presence of SOC, the pairing wave function develops an odd-parity spin-triplet correlation coexistent with the even parity spin-singlet component\cite{Edelstein_triplet}. Such a triplet correlation at the Nb-Pt interface has also been experimentally detected via low-energy Muon spin resonance experiments \cite{NB/Pt_paramagnetic_interface}. Secondly, it has been predicted\cite{spinhallinJJ} that in current biased SNS junctions hosting spin-orbit interaction in the barrier, the spin triplet component of the pairing function leads to an accumulation of spin moment analogous to the normal spin-Hall effect\cite{SHE_Pt3,SHE_Pt2,SHE_Pt,SHE_Pt1}. It is, however, important to note that no spin-current is expected in this case\cite{spinhallinJJ}, unlike the normal spin-Hall effect. As the general premise of this prediction is not restrictive to Rashba or Dresselhaus type SOC, it is reasonable to expect a similar effect in our vertical Nb-Pt-Nb junction geometry with strong intrinsic spin-orbit coupling in the Pt layer. Once we establish the presence of an effective spin moment ($M_S$) in the junction, it is immediately recognizable in the Figure 1 (b) that, depending on the direction of the external magnetic field, the phase generated by the non-equilibrium spin-moment $M_S$ in the Josephson barrier can either add or subtract to the current-phase relation as $I_c = I_{\text{c0}}\sin(\phi_H \pm \phi_M)$. Such a non-reciprocal phase addition naturally leads to an asymmetry in Josephson current. Here $\phi_H$ and $\phi_M$ are the phases corresponding to the external magnetic field and the spin moment $M_S$, respectively. In the next section we experimentally demonstrate the Josephson diode effect in these trilayer Nb-Pt-Nb vertical nano-junctions.

\section*{Observation of Josephson diode effect in Nb-Pt-Nb nano-pillar junctions}
\begin{figure}[h!]
    \centering
    \includegraphics[width=1\textwidth]{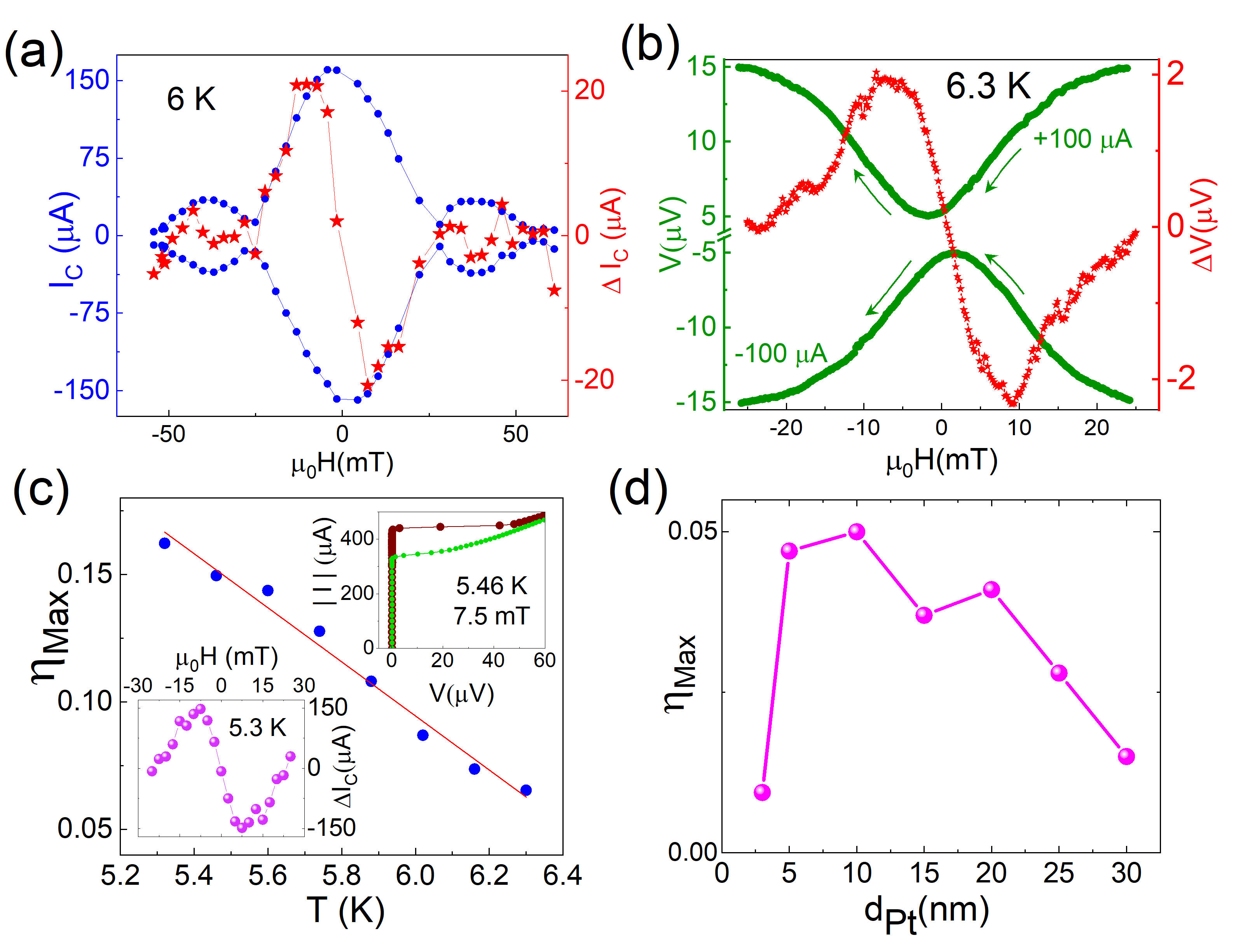} 
    \caption{\textbf{Observation of Josephson diode effect and its dependence on temperature and d$_\text{Pt}$.}(a) Critical current Fraunhofer pattern of a Nb-Pt(20nm)-Nb nano junction extracted from current-voltage curves measured at various magnetic fields is plotted in the left hand side axis. The right hand side axis plots the difference in magnitude of the positive and negative critical currents, showing a clear field dependent diode effect. (b)  Voltage across the same junction measured as a function of an in-plane magnetic field is compared in the left hand side axis, for +100$\mu$Amp and -100$\mu$Amp bias currents. The difference $\Delta V = \lvert V^{+} \rvert - \lvert V^{-} \rvert$ between the absolute values of the voltages measured under positive and negative current biases as a function of $H$ is plotted in the right hand side axis, which also represents field dependent non-reciprocity in the Josephson critical current, equivalent to the data in the first panel. (c) The maximum diode efficiency plotted as a function of temperature for the same junction. The top inset in this figure shows the current-voltage curve at 5.46 and 7.5 mT field, highlighting the asymmetry between the positive and negative Josephson currents. The bottom inset shows the field dependence of the difference in critical current of the same junction at 5.3 K. (d) The maximum diode efficiencies are plotted as a function of the Pt barrier thicknesses, measured at 0.9 T$_{JJ}$ for the respective nano junctions.}

    \label{Fig.2}
\end{figure}
The Josephson diode effect (JDE) has been extensively studied in junctions with Rashba spin–orbit coupling (SOC) under an applied magnetic field. In such inversion asymmetric systems, broken time reversal symmetry leads to a finite Cooper-pair momentum, as the Fermi contours shift due to Zeeman coupling. This results in an asymmetric current–phase relation, giving rise to nonreciprocity in the critical current\cite{RSOCdiodeAlAs-InAsplanar,RSOCdiodeBergerate,Rsocfinitemomentum1}. As described in the previous section, an asymmetry in the current-phase relation of vertical JJs can also be generated by supercurrent induced spin moment in the barrier. Since the relative orientation of the external field and the induced spin-moment defines the phase asymmetry (hence current asymmetry) in this geometry, no Josephson diode effect may be expected at zero external magnetic field. Figure 2(a) shows the characteristic critical current Fraunhofer pattern of a Nb-Pt(20 nm)-Nb nano-junction measured at 6 K. This was extracted from the current-voltage measurements performed with in-plane magnetic field. The Fraunhofer pattern establishes Josephson coupling and the quality of the junction. However, a clear, field dependent asymmetry is observed between the positive and negative critical currents, especially within the first lobe of the Fraunhofer pattern. The difference $\Delta I_C$($I_c^{+}-I_c^{-}$) is plotted on the right hand side axis on the same graph which clearly demonstrates the Josephson diode effect. Figure 2(b) shows the V(H) curves (corresponding to Fraunhofer patters) of the same junction at 6.3 K, measured with $+100~\mu\text{A}$ and $-100~\mu\text{A}$ bias currents. The resistance minima in the two cases, corresponding to the maximum Josephson current through the junction, exhibit a clear relative shift of magnitude $\sim$3.3 mT. Throughout this study the magnetic field was always swept from positive to negative field, although no hysteresis in V(H) was observed between forward and reverse sweeps, as shown in supplementary Figure S4. The shift in the V(H) pattern upon bias current reversal was a consistent feature of over 15 devices measured across the entire series of samples with varied Pt thickness. The magnitude of the shift in V(H), however, was dependent on the thickness of the Pt barrier. Some representative V(H) data for other thickness of Pt are presented in the supplementary Figure S3. Such a shift in V(H) (corresponding to a shift in I$_C$(H)) is a direct signature of the Josephson diode effect, also shown in earlier studies\cite{Turini2022}. The difference ($\Delta$V), between the V(H) curves corresponding to the +ve and -ve bias currents, quantifies the magnetic field dependence of the degree of non-reciprocity. As plotted on the right hand side axis in the Fig 2(b), JDE in this junction appears to be most prominent around $\sim\pm$ 7.5 mT with a change in sign across B=0. Therefore, in good agreement with the direct critical current measurements, the voltage across the junction also confirms non-reciprocal behavior on the function.

In order to examine the temperature dependence of the JDE, we measured the $I-V$ characteristics of this junction at several temperatures under a magnetic field of 7.5 mT (corresponding to the maximum diode effect for this junction) and estimated the diode efficiency using the relation $\eta = \frac{I_c^{+} - I_c^{-}}{I_c^{+} + I_c^{-}}$\cite{unisersalJDE}
which is shown in the Fig. 2(c). A near linear increase in non-reciprocity was observed at lower temperatures as seen from the linear fit in this figure. In the bottom inset of Fig 2(c) we plot the $\Delta I_C$($I_c^{+}-I_c^{-}$) at 5.3 K, which shows an absolute current asymmetry of as large as 150 $\mu$A at the highest point. A representative current-voltage curve showing maximum asymmetry in the critical current at 5.46K has been plotted in the top inset of Fig 2(c). The maximum value of diode efficiency as a function of Pt thickness($d_\textit{Pt}$) reveals a non-monotonic trend in Fig 2(d), maximizing around 10-15 nm of Pt. In order to standardize the dataset, this curve was extracted from direct measurements of $I_c^{+}$ and $I_c^{-}$ at 90\% of $T_{JJ}$ at magnetic fields corresponding to the maximum diode effect for each Pt thickness. Therefore, the data in Fig. 1 firmly established the Josephson diode effect across the parameter space of temperature, magnetic field, and thickness of the Pt barrier. Since the diode effect in the Nb-Pt-Nb Josephson junctions hinges on the generation of a spin moment (M$_S$) in the Pt layer, in the following section we experimentally demonstrate the predicted\cite{spinhallinJJ,Hikino_tripletJJ} supercurrent induced spin magnetization in these junctions.

\section*{Supercurrent spin Hall effect in Nb-Pt-Nb nano-junctions}
 If a spin moment is indeed generated in the Pt layer of the junction then, in the current biased state, the barrier can effectively behave as a pseudo-magnetic layer. We follow two independent strategies to establish this effect. Firstly, we recognize that the spin moment is generated by the spin-polarized (spin-triplet) component \cite{spinhallinJJ, Hikino_tripletJJ} of the pairing function in the Pt layer which constitutes only one fraction of the supercurrent across the junction. The other fraction of supercurrent across the junction is carried by the unpolarized spin-singlet component. Therefore, while traversing the diffusive barrier in the current carrying state, the spin singlet Cooper pairs can sense the induced spin moment in the Pt layer and gain a finite momentum. In that case, very similar to ferromagnetic JJs, some signature of $0-\pi$ type transition in Josephson coupling can be expected. In a second approach we argue that, if a spin moment is indeed established in the Pt layer due to the proximity of superconducting Nb layer then, placing a ferromagentic (Ni) layer next the Pt layer would lead to a spin-valve like magnetoresistance. While the orientation of the magnetic moment in Ni (M$_{Ni}$) can be tuned by an external magnetic field, the orientation of the spin moment (M$_{S}$) can be independently tuned by the bias current. Here the parallel orientation of M$_{S}$ and M$_{Ni}$ would cause less resistance in the Pt-Ni bilayer compared to the anti-parallel orientation, when the Nb electrodes are superconducting.  In the following sub sections we discuss these two independent measurements in detail.

\subsection*{Signature of $0-\pi$ transition in Nb-Pt-Nb nano-junctions due to SSHE}
In the case of ferromagnetic Josephson junctions and multilayers \cite{SFSCuNi,0-piCuNi,SFSjunction,SFS0-pithesis, Jiang_oscTc, Lazar_oscTc}, an oscillatory thickness dependence of critical parameters such as I$_C$ or T$_C$ has been widely reported. This effect is a consequence of the finite momentum ($\Delta p$) gained by the singlet Cooper pairs upon entering the magnetic barrier following $\Delta p \simeq \frac{2E_{Ex}}{\hbar v_F} $, where $E_{ex}$ is the ferromagnetic exchange energy, and $v_F$ is the Fermi velocity. In the present case of fully non-magnetic Nb-Pt-Nb junctions, spin-singlet Cooper pairs can gain a momentum only via an induced spin moment in the barrier layer. In order to look for an oscillatory trend in the junctions transition temperature, we measured a series of Nb-Pt-Nb junctions with varying thickness of Pt layer. Figure 3(a) presents the temperature dependence of resistance of a Nb-Pt-Nb nano-pillar junction with Pt thickness of $\sim$ 30 mn. In this figure the superconducting transitions of the junction electrodes is marked as T$_{SC}$, and a lower temperature transition of the junction is marked as T$_{JJ}$. Multiple transitions seen near T$_{SC}$ arise from variations in the width of the Nb electrodes introduced during the focused ion beam fabrication process. This is the general feature of the R(T) curves for all the junctions fabricated from the series of trilayers with Pt thickness varying from 3 nm to 30 nm [shown in supplementary Figure S2]. The inset in Fig 3(a) shows the $I$–$V$ characteristics of the same junction at 3.4K. In Fig 3(b) we plot the T$_{JJ}$ for all measured devices, normalized to the respective T$_{SC}$. Note that multiple devices were measured corresponding to each thickness. The non-monotonic variation of the normalized junction transition temperature as a function of Pt barrier thickness, akin to the case of ferromagnetic JJs\cite{Jiang_oscTc, Lazar_oscTc,0-pifitRobinson}, is apparent from this figure. Considering the non-magnetic Pt barrier in this case, the observed oscillatory behavior of $T_\text{JJ}/T_\text{SC}$ indicates the presence of an internal spin magnetization which is experienced by the singlet Cooper pairs in the junction. An expression of the type $e^{-d / \xi_{1}} 
\sin\!\left( \frac{d - d_0}{\xi_{2}} \right)$, is generally used to fit the thickness dependence of critical current for diffusive SFS junctions \cite{glic2017critical}. Here 'd' is the barrier thickness,$d_0$ is the thickness of barrier corresponding to the first minimum in critical current,  $\xi_{1}=35\,\mathrm{nm}$\cite{PtXi1} is coherence length of induced superconducting order in the barrier, $\xi_{2}$ is given by $\sqrt{\frac{\hbar D}{E_{ex}}}$. Emphasizing the similar effect of a spin moment in Pt barrier in these junctions as the magnetic moment in a typical SFS junction, from the perspective of the singlet Cooper pairs traversing the diffusive barrier, we use the same expression to simulate the thickness dependence of $T_{JJ}$ in Fig 3(b). The qualitative dependence of the simulated I$_C$R$_N$ curve, plotted as a solid line on the right hand side axis of the Fig 3(b), matches well with data for an effective exchange energy of 14.8 meV. For this simulated curve the diffusion coefficient D was taken as $D_F = \frac{1}{3} v_F l = 2.04 \times 10^{-3}\,\mathrm{m}^2/\mathrm{s}$, using $v_F = 1.46 \times 10^6\,\mathrm{m/s}$ and a mean free path $l = 4.19\,\mathrm{nm}$ for Pt thin films grown in the same sputter deposition system~\cite{PtfermiV}. We must, however, clarify here that Pt layer does not host any ferromagnetic exchange mechanism and hence it is not expected to show any magnetic hysteresis. Consequently, magnetic field dependence of the voltage across the junctions (V(H) curve), which is essentially the manifestation of the critical current Fraunhofer patterns\cite{Tapas_JJ}, does not show hysteresis between forward and reverse field sweeps[shown in Supplementary Fig S4]. We note here that we have ruled out any magnetic impurities in the junction barrier by performing energy dispersive X-ray spectroscopy with long acquisition time, as shown in the supplementary Fig S6. 

\begin{figure}
    \centering
    \includegraphics[width=1\textwidth]{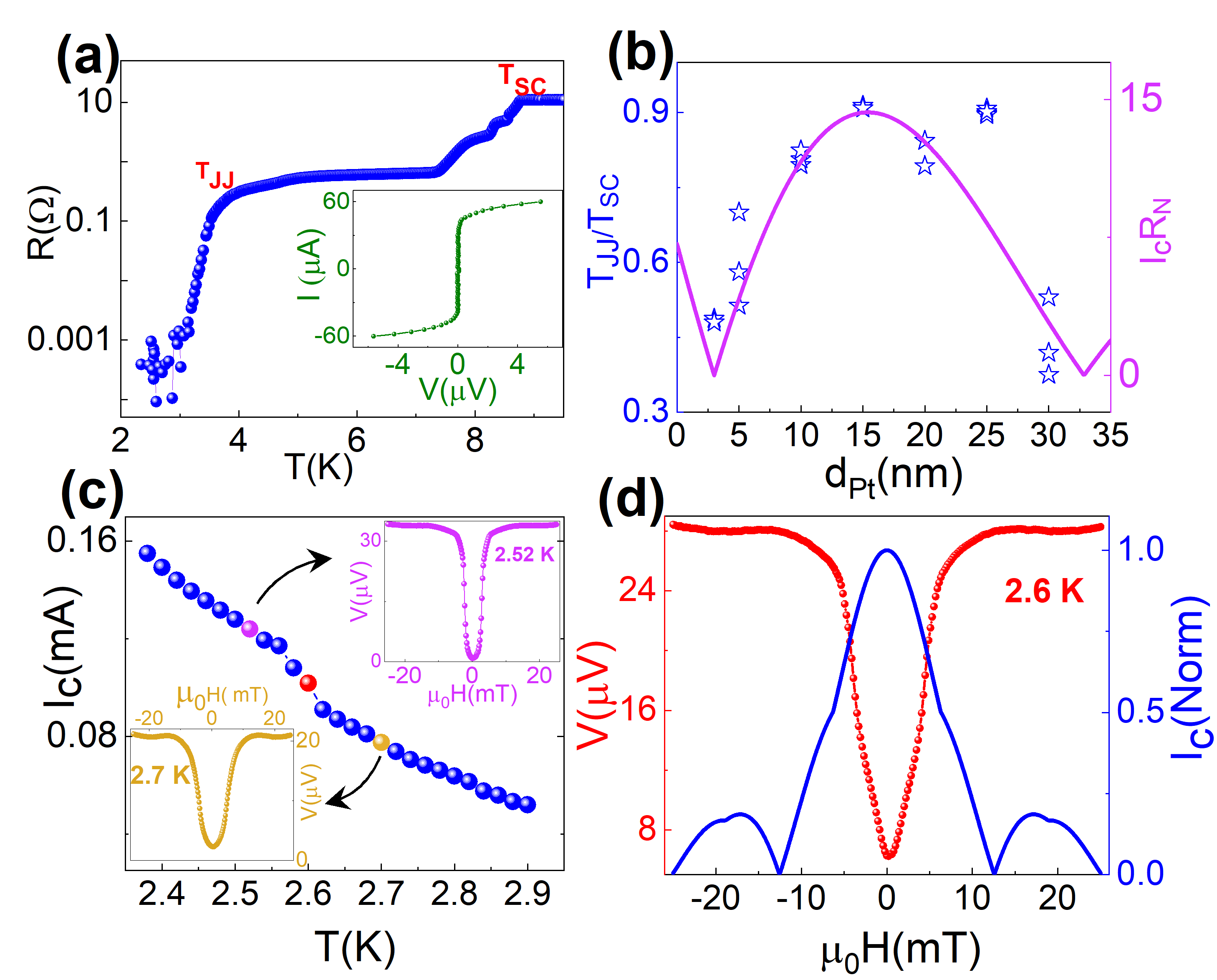} 
    \caption{\textbf{Signature of $0$–$\pi$ transition as a consequence of a spin moment in the Nb-Pt-Nb JJs)}(a) Resistance as a function of temperature at zero field for a Josephson nano device with $d_\text{Pt} = 30$\,nm, showing the superconducting transitions of the electrode at $T_\text{Sc}$ and proximatization transition $T_\text{JJ}$.\textit{Inset:} I–V characteristics of the same JJ at zero magnetic field at 3.4K. No asymmetry is noted in this IV curve due to the zero field condition. (b)The junction transition temperature, normalized with the electrode transition temperature ($T_\text{JJ}/T_\text{Sc}$) is plotted as a function of Pt thickness in the left hand side axis. Multiple junctions were measured at each thickness. The solid line, plotted on the right hand side axis, represents a calculated dependence of critical current on the barrier thickness for a SFS junction, as discussed in the text. (c) The temperature dependence of critical current $I_c$ of the Nb-Pt(30\,nm)-Nb JJ is plotted as a function of temperature. The insets show the V(H) curves measured at 2.52 K and 2.7 K. (d) V(H) curve of the same junction measured at 2.6 K, showing signature of second harmonic Josephson current component. The Solid line is a simulated critical current Fraunhofer consisting of first and second harmonic current components at 4:1 ratio. }
    \label{}
\end{figure}

To further substantiate the presence of a subtle $0-\pi$-like transition in these non-magnetic junctions, attributed to an induced spin moment, we investigated the temperature dependence of the critical current in a Nb–Pt–Nb junction with a 30 nm Pt barrier (Fig. 3(c)). A sudden change in the slope of the $I_CR_N$ product was observed close to 2.6 K. It is known that near the $0-\pi$ transition, the first-harmonic contribution to Josephson current is somewhat suppressed. As a result, signatures of the second harmonic component of Josephson current emerge, which is otherwise overwhelmed by the first harmonic component. Therefore, the emergence of a second-harmonic component in the Josephson coupling is considered to be a clear signature of the $0-\pi$ transition point in Josephson junctions \cite{frolovSFS,harlinger2ndharmonicSFS,Buzdin2ndharminicSFS,wangsecondharmonicSFS}. To verify if the change in the slope of the critical current at 2.6 K in the Fig 3(c) corresponds to a $0-\pi$ transition point, we closely examined the V(H) curves of this junction measured around 2.6K. The V(H) curves measured at 2.52K and 2.7 K, plotted as insets of the Fig 3(c), show a regular pattern. However, the V(H) curve measured at 2.6 K (plotted on the left hand side axis Fig 3(d)) shows a bump in the primary lobe, which is a typical signature of the presence higher harmonic Josephson coupling. The solid line in the Fig 3(d) is a simulated Fraunhofer pattern comprising of the first and second harmonic components as $I_c = I_{c1} \left| \frac{\sin(aH)}{aH} \right|+ I_{c2} \left| \frac{\sin(2 aH)}{2 aH} \right|$. Here, H is the applied field, 'a' is a geometric constant and $I_{c1}$ and $I_{c2}$ are the first and second harmonic critical currents, respectively. The simulated curve in Fig 3(d) corresponds to $\frac{I_{c2}}{I_{c1}}\sim\frac{1}{4}$. The signature of second harmonic component of Josephson current was also apparent in the V(H) curves of the junctions with Pt thickness close to the cusp in Fig 3(b), such as  Pt 3 nm and 5 nm, as shown in the supplementary Fig S3. The signature of $0-\pi$ transition in these non-magnetic junctions is a strong indicator of the presence of a spin moment in the Josephson barrier.

\subsection*{Emergent spin-valve effect in Nb-Ni-Pt-Nb nano-junctions due to SSHE}
\begin{figure}[h!]
    \centering
    \includegraphics[width=1\textwidth]{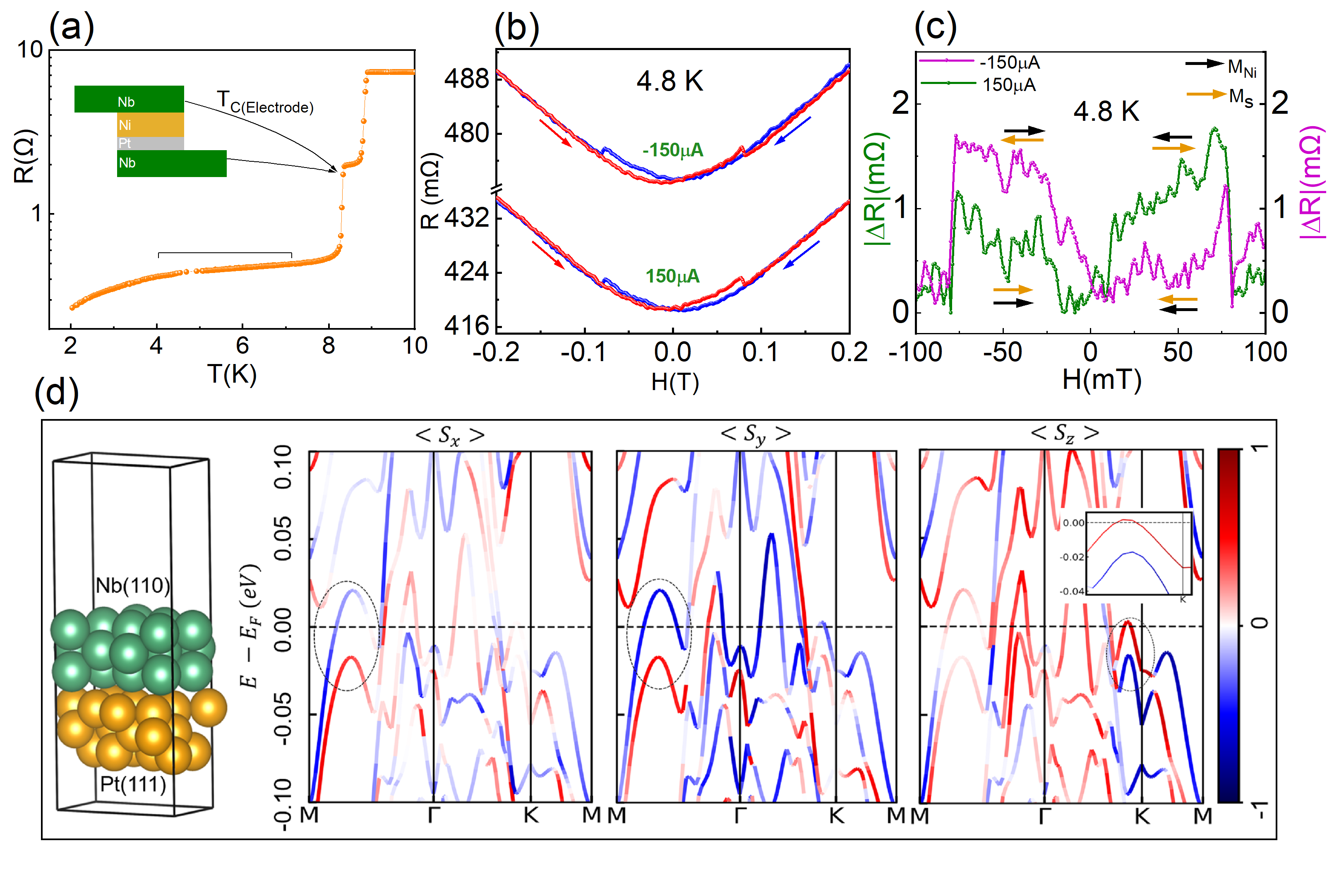} 
    \caption{\textbf{Spin valve like behavior in Nb/Pt/Ni/Nb junctions and spin polarized band structure at Nb-Pt interface}(a) Resistance ($R$) as a function of temperature ($T$), showing the superconducting transition of the wide track at $8.9~\text{K}$ and the junction electrode at $8.3~\text{K}$. (b) Resistance as a function of the applied magnetic field measured at $4.8~\text{K}$ are plotted for bias currents of $+150~\mu\text{A}$ and $-150~\mu\text{A}$, in forward and backward magnetic field sweep conditions. (c) The magnitude of the resistance difference $|\Delta R| = |R(H_{\text{forward}}) - R(H_{\text{backward}})|$, extracted from Fig 4(b) is plotted as a function of magnetic field. The relative orientations of the induced spin moment $M_{\textit{S}}$ and the Ni moment $M_{\textit{Ni}}$ are illustrated in the various regimes of the data. Notice that for a fixed magnetic field (fixed M$_{Ni}$) reversal of bias current leads to the switching of M$_S$, leading to two different resistance states, like a spin valve. (d) Constructed Nb(110)/Pt(111) heterostructure model used for first-principles calculations, with Nb layers stacked on Pt along the $z$-direction and the resultant spin-resolved band structure heterostructure. The band dispersion along the high-symmetry path M--$\Gamma$-K-M is shown with spin projections $\langle S_x \rangle$, $\langle S_y \rangle$, and $\langle S_z \rangle$. The color scale represents the expectation values of the corresponding spin components, as indicated by the color bar. The Fermi level is set to zero energy. Regions exhibiting pronounced spin splitting are highlighted by dotted circles. The inset in the $\langle S_z \rangle$ panel shows a magnified view of the spin splitting near the K point. }
    \label{fig:mygraph}
\end{figure}
The second approach to establish the existence of a spin moment in the Pt layer in proximity of a superconducting Nb layer relies on realizing a spin-valve like configuration with a ferromagnetic layer. For this experiment we fabricated a Nb(160nm)/Pt(5nm)/Ni(20nm)/Nb(160nm) nano pillar device following the same fabrication route as the Nb-Pt-Nb devices. The inset of  Fig 4(a) shows a schematic of the device configuration. The temperature dependence of resistance of this device is shown in the Fig 4(a) in log scale. Sharp drops in resistance of the junction close to 8.7 K and 8.3 K corresponds to the superconducting transition of the Nb tracks and the narrow Nb electrodes, respectively. Below 4 K a further gradual drop in the resistance indicates the onset of proximatization of the thick Ni layer. Therefore, in the range of 4 K to 7 K (marked by a bracket in Fig 4(a)), the junction resistance predominantly comes from the Ni-Pt bilayer only. We measured the in-plane magnetoresistance of the junction in this temperature range. Figure 4(b) shows a representative MR data measured at 4.8 K with 150 $\mu$A and -150 $\mu$A bias currents. The plots are shifted along the y-axis for clarity. For +150 $\mu$A bias current we notice a slight higher magnetoresistance for the forward field sweep case compared to the reverse field sweep. The opposite trend was observed for -150$\mu$A bias current. In order to highlight this effect, we have plotted difference ($|\Delta R|$) between the forward sweep and reverse sweep resistances in the Fig 4(c), for both the bias currents. Clearly, for opposite bias currents, the $|\Delta R|$ is higher on the opposite side of zero field. Similar to the case of multilayer spin-valves\cite{spinvalve1,Spinvalve2}, we can map the higher resistance to a relative opposite orientation of Ni moment ($M_{Ni}$) to the induced spin moment $M_S$. The external magnetic field determines the direction of the Ni moment, whereas the direction of spin moment depends on the bias current direction. Therefore, for a fixed direction of external magnetic field, the device switches between a higher or lower resistance state depending on the direction of bias current. This observation is another strong evidence for the presence of spin magnetization $M_S$ in the Pt layer, due to the triplet correlation generated in Pt by the proximity of the superconducting Nb layer.

\section*{Discussion}
Unlike the most widely studied diode systems, relying on the Rashba SOC, in our trilayer structure the origin of the in-field JDE can be linked to the intrinsic spin-orbit coupling effect in the Pt layer. As such, the energy spectrum of centrosymmetric Pt is spin degenerate. However, the structural inversion symmetry is naturally broken at the Nb-Pt interface. To understand the microscopic origin of spin splitting at the Nb-Pt interface, arising from interfacial inversion symmetry breaking and the strong SOC of Pt, and its role in enabling spin-triplet correlations, we performed first-principles calculations. We find that in the absence of SOC all electronic bands are spin degenerate, reflecting the preservation of spin-rotational symmetry, as shown in supplementary Figure S8. Upon inclusion of SOC, pronounced momentum-dependent spin splittings (0.3-0.5 meV) emerge near the Fermi level, as illustrated in the calculated spin projected band structures plotted along both in-plane (S$_x$ and S$_y$) and out-of-plane (S$_z$) spin components, in the Figure 4(d). Further details of the calculations are given in the supplementary information. Fig 4(d) indicates that SOC induces significant spin–orbital mixing and generates momentum-dependent spin textures in the band structure. We must, however, mention that even though significant spin splittings were observed close to the Fermi level, the net spin moment added up to zero, as expected for the non-magnetic Nb-Pt system. Significant spin splitting at the Fermi level facilitates spin mixing at the Nb-Pt interface, thereby allowing the generation of an odd-parity spin-triplet component in the Cooper pair wavefunction\cite{GorkovSingletTriplet}, when the Nb layer becomes superconducting. The spin moment arising out of the spin-triplet component of the pairing function\cite{spinhallinJJ,Hikino_tripletJJ} ultimately drives the non-reciprocity in the vertical Nb-Pt-Nb junctions. With increasing magnitude of critical current the net induced spin moment can also increase, leading to an enhanced diode efficiency. Therefore, the observed monotonic increase in the diode efficiency (shown in Fig 2(c)) at lower temperatures directly correlates with increasing critical current of the junction. Similarly, the non-monotonic dependence of diode efficiency on the thickness of the Pt barrier (Fig 2(d)), correlates well with the non-monotonic behavior of normalized junction transition temperature ($\frac{T_{JJ}}{T_{SC}}$) shown in the Fig 3(b). This observation also reaffirms the direct dependence of the magnitude of the induced spin-moment (and hence diode efficiency) on the supercurrent, similar to the dependence of normal spin-Hall effect on the magnitude of bias current. The monotonically increasing diode efficiency at low temperatures also indicates that the emergent spin-polarization can not be linked to a quasiparticle driven normal spin-Hall effect in the Pt barrier. This is because the thermally excited quasiparticle density dies out exponentially as \(e^{-\Delta /k_{B}T}\) at lower temperatures\cite{tinkham1996}, which is a trend just opposite to the observed temperature dependence of the JDE efficiency.  

The existence of a true spin moment led to a spin-valve effect in Nb-Pt-Ni-Nb junction (Fig 4) and also resulted in $0-\pi$ transition in Nb-Pt-Nb junctions (Fig 3). Theoretically, a $0-\pi$ like transition has also been predicted \cite{SOC0-pi} in planar JJs with 2DEG barrier as a function barrier length. A pseudo-magnetic effect due to an in-plane current in the 2DEG with strong Rashba SOC was argued to be the origin of $0-\pi$ oscillations in such systems\cite{SOC0-pi,Bezuglyi_ZeemanSOC}. By contrast, the direction of current in our device geometry was parallel to the interfacial symmetry-breaking field, which excludes any possible role of a pseudo-magnetic field in Pt barrier.  A $0-\pi$ like transition in current-phase relation was reported to lead to in-plane magnetic field dependent reversal in the sign of the diode effect in JJs with InAs/InGaAs quantum well barriers\cite{costa0-pidiode}. This was described as a result of the integrated response of multiple Andreev transport channels, which acquire different phases from the external magnetic field. By nature this observation in the ballistic regime is distinct from the barrier length and temperature dependent $0-\pi$-like oscillations in our Nb-Pt-Nb diffusive junctions. 

\section*{Conclusion}
In conclusion, we demonstrated Josephson diode effect above liquid Helium temperature, in nanoscale vertical junctions fabricated from simple Nb-Pt-Nb trilayers. The observation was possible to describe by the existence of a spin-moment in the junction barriers which, in the presence of an external magnetic field, tunes the current phase relation in a non-reciprocal manner leading to the observed diode effect. Band structure calculations showed momentum dependent spin splitting at Nb-Pt interface. This was conducive for inciting the spin-polarized triplet component of the pairing function, crucial for the generation of the spin moment. The existence of the spin moment due to the supercurrent spin-Hall effect was substantiated for the first time via the observation of a $0-\pi$ transition in Nb-Pt-Nb junctions and a spin valve effect in Nb-Ni-Pt-Nb junctions. The importance of our finding lies in the fact that the fully metallic architecture of Nb–Pt–Nb Josephson diodes makes them readily integrable into complex, scalable superconducting electronics.

\section*{Methods}
\subsection*{Multilayer deposition}
A series of trilayer Nb/Pt/Nb films were deposited on cleaned $5 \times 5$\,mm$^2$ Si/SiO$_2$ substrates using DC magnetron sputtering from high-purity Nb and Pt targets (99.998\%) at a base pressure of $6 \times 10^{-8}$\,mbar~\cite{Tapas_JJ}. Prior to deposition, the sputtering chamber was baked for 12\,h and subsequently cooled for 12\,h to minimize residual moisture and achieve a stable base pressure. A residual gas analyzer (RGA) was used to monitor the partial pressure of water vapor, and a titanium sublimation pump (TSP) was employed immediately before deposition to further reduce the water content in the chamber.

In all samples, the bottom and top Nb layers were kept fixed at a thickness of 160\,nm, while the Pt barrier thickness was varied from 3\,nm to 30\,nm across the series. The Pt thickness was controlled by adjusting the rotation speed of the substrate holder during deposition. The Pt layer thickness was calibrated using X-ray reflectivity (XRR) measurements by fitting the Kiessig fringe periodicity with simulations performed using the \textsc{GenX} package, as shown in the Supplementary Information. Prior to device fabrication, 2$\mu$m-wide tracks of the trilayer samples were prepared by depositing films on lithographically patterned substrates followed by a lift-off process.
\subsection*{Junction fabrication and measurements}
The nano-patterning of the Josephson junctions was carried out using a focused ion beam (FIB) technique in a Zeiss Crossbeam 340 system. The process involved three sequential stages on lithographically pre-patterned tracks. A custom-designed stub with a 54\textdegree inclined face was employed to mount the sample, allowing the required angular alignment between the sample surface and the FIB beam for milling from both the top and side directions. 

In the first stage, an ion current of 100\,pA at 30\,kV was applied to narrow the 2$\mu$m track width to approximately 500\,nm. During the second stage, the current was reduced to 10\,pA at the same accelerating voltage to further decrease the width to about 200\,nm and to polish the sidewalls, thereby ensuring a clean and well-defined junction interface. These steps were performed with the ion beam oriented perpendicular to the sample surface. For the final stage, the sample was tilted such that the angle between the sample normal and the FIB beam was about 87\textdegree. A low beam current of 5\,pA at 30\,kV was then used to mill two narrow slots, each roughly 60\,nm wide, defining a nano-pillar with a rectangular cross-section.

Electrical transport measurements on the Josephson junction (JJ) devices were carried out down to 2K, in a closed cycle cryostat from Cryogenic limited, equipped with a high precision low current magnet power supply. Devices were always zero field cooled to avoid any flux trapping effects. Electrical measurements were performed in the standard four probe geometry where all signal lines were filtered through low pass grounded RC filters with cutoff frequency of 1 KHz. In-field measurements were performed with magnetic field along the plane of the junction, in all cases. Remanent magnetic field of the system was subtracted from all $V(H)$ data. 

\subsection*{Electronic Structure Calculations}
First-principles calculations of the electronic band structure of the Nb/Pt 
interfacial heterostructure were carried out within Density Functional Theory (DFT)~\cite{Hohenberg1964,Kohn1965} using the Vienna \textit{Ab initio} Simulation Package (VASP)~\cite{kresse1996efficiency}. The interaction between ionic cores and valence electrons was described within the projector augmented-wave (PAW) framework~\cite{kresse1996efficient,kresse1999ultrasoft}, while exchange–
correlation effects were treated using the Perdew–Burke–Ernzerhof generalized-gradient approximation (GGA)~\cite{perdew1996generalized}. The Kohn–Sham states were expanded in a plane-wave basis with a kinetic-energy cutoff of 480 eV. To capture the inversion-symmetry-broken electronic environment of the Nb/Pt interface, slab geometries separated by a vacuum region exceeding 20~\AA\ were constructed along the out-of-plane direction. Brillouin-zone sampling employed $\Gamma$-centered Monkhorst–Pack meshes of $12 \times 12 \times 1$ for slab calculations and $18 \times 18 \times 18$ for bulk reference calculations. Structural relaxations were performed until the total-energy and Hellmann–Feynman force convergences reached $10^{-6}$ eV and $10^{-3}$ eV/\AA, respectively. During optimization, the in-plane lattice parameters were fixed to simulate epitaxial growth conditions, whereas all internal atomic coordinates were fully relaxed.\\

Spin–orbit coupling (SOC) was subsequently included self-consistently within the non-collinear formalism using the second-variational approach, with crystalline symmetry disabled to accurately resolve interfacial spin splitting and momentum-dependent spin textures. The calculations reveal pronounced SOC-driven splitting near the Fermi level originating from the combined effects of strong Pt spin–orbit interaction and broken inversion symmetry at the Nb/Pt interface. 
Electronic dispersions were evaluated along the high-symmetry directions of the surface Brillouin zone. Additional computational details and spin-texture analyses are provided in the Supplementary Information.

\bmhead{Data Availability}
The data that support the findings of this study are available within the Article and its Supplementary Information. Further information is available from the corresponding authors on reasonable request.

\putbib[main]   

\end{bibunit}

\bmhead{Acknowledgments}
Authors are grateful for financial support from NISER through plan project RIN-4001 and Department of Science and Technology through project no: ANRF/ARG/2025/007445/PS. Authors also acknowledge discussions with Dr. Kush Saha and Dr Ajaya K Nayak from NISER Bhubaneswar, for useful discussions and suggestions.  

\bmhead{Author Contribution}
DN performed the device fabrications and measurements. DN and KS analyzed the data and wrote the manuscript. DR and PS performed the band structure calculations and compiled the relevant text. KS supervised the project. 

\bmhead{Declaration}
The authors declare no competing interests.

\newpage
\clearpage
\begin{centering}
    \section*{Supplementary Information}
\end{centering}

\begin{bibunit}
\setcounter{figure}{0}
\setcounter{table}{0}

\renewcommand{\thefigure}{S\arabic{figure}}
\renewcommand{\thetable}{S\arabic{table}}

\makeatletter
\renewcommand{\fnum@figure}{\textbf{Supplementary Fig.~\thefigure}}
\renewcommand{\fnum@table}{\textbf{Supplementary Table~\thetable}}
\makeatother

\begin{table}[ht]
\centering
\caption{A collection of reported Josephson diode performances}
\label{}
\begin{tabular}{c c c c p{6cm}}
\hline\hline
Josephson Junction & Geometry & $\eta\% ( \frac{I_c^{+} - I_c^{-}}{I_c^{+} + I_c^{-}}\times 100)$&T & Mechanism of Field-Induced JDE \\ 
\hline
Al/InGaAs/Al\cite{SJDE1} &Planar  &8\% & 30mK  & In-plane field–induced finite-momentum pairing due to Rashba SOC \\
Al/InAs nanosheet/Al\cite{SJDE2} & Planar & 3\% & 30mK& Gate-tunable Rashba SOC under an in-plane field leads to tunable JDE.   \\
Al/InSb nanowire/Al\cite{SJDE3} & Planar & 8\% & $<$1.8K
 & Finite-momentum Andreev bound states induced by SOI and Zeeman fields, with multi-gate tuning to maximize diode efficiency. \\
Al/InAs/Al \cite{SJDE4}& Planar & 24\% & 14mK & Finite-momentum Cooper pairing enhanced by phase tuning in a three-terminal geometry. \\
$\mathrm{NbSe}_2/\mathrm{NbSe}_2$ \cite{SJDE5} &Vertical  & 10.4\% & 30mK & Nonsinusoidal CPR combined with interfacial SOI inducing finite-momentum pairing in a weak-coupling regime. \\
Nb/InAs nanoflag/Nb\cite{SJDE6} &Planar &$\sim$2.5\% & 30mK  & Finite-momentum pairing induced by Rashba SOC and in-plane magnetic field; JDE efficiency is not enhanced by gate biasing \\
Al/InAs/Al\cite{SJDE7} & Planar  & 6\% & 30mK & Modified Andreev reflection via electrostatic gating and SOC reshapes the Current-Phase Relation to induce diode behavior. \\
$\mathrm{Al}/\mathrm{Cd}_3\mathrm{As}_2/\mathrm{Al}$\cite{SJDE8} & Planar &20\% & 12mK  & Self-field induced JDE in planar geometry, where out-of-plane field and induced self-field jointly break mirror and time-reversal symmetries. \\
Al/InAs nanowire/Al\cite{SJDE9} &Planar  & $\sim$3.5\% & 10mK &Non-local symmetry breaking (TRS, Inversion) enables JDE, with efficiency controlled by tuning non-local phase and electrostatic gating.  \\
Nb-Pt-Nb (This work ) & Vertical & $\sim$ 17\% & 5.3 K & Intrinsic SOC in Pt barrier led to superconducting spin-Hall effect and a net spin moment in the barrier, which caused the non-reciprocal response in the presence of magnetic field.   \\

\hline\hline
\end{tabular}
\end{table}

\subsection*{Note 1: Thickness and roughness of Pt barrier films}
The thickness of the Pt barrier over the series of samples was controlled via in-situ rotation speed of the substrate holder under the sputtering source. Fig S1 shows the glancing angle X-ray reflectivities of three different thicknesses of standalone Pt films in the range used in the series of Nb-Pt-Nb trilayers. X-ray reflectivity (XRR) measurements were simulated using the \textsc{GenX} software for Pt films deposited at different substrate holder rotation speeds of $3^{\circ}\,\mathrm{s^{-1}}$, $2^{\circ}\,\mathrm{s^{-1}}$, and $1.2^{\circ}\,\mathrm{s^{-1}}$. The experimentally observed Kiessig oscillations show excellent agreement with the simulated reflectivity curves, confirming the reliability of the fitting. A systematic and nearly linear increase in the Pt film thickness is observed with decreasing rotation speed, demonstrating precise control over the deposition rate. The interface roughness extracted from the XRR fitting remains below $0.3\,\mathrm{nm}$ for all rotation speeds, indicating smooth and high-quality Pt films.
\begin{figure}[h!]
    \centering
    \includegraphics[width=.7\textwidth]{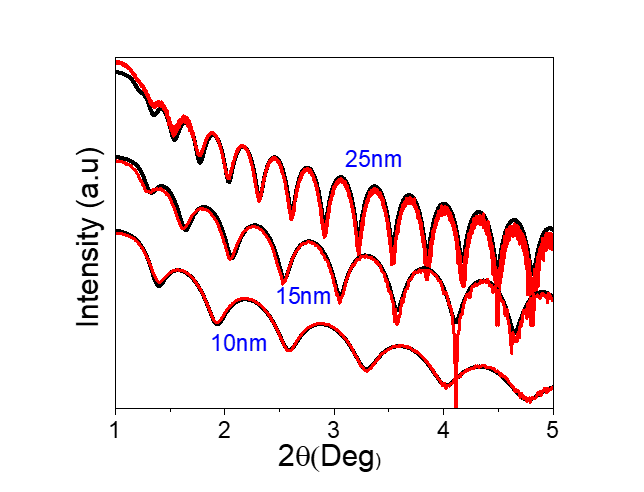} 
    \caption{ X-ray reflectivity measurements performed on Pt thin films to estimate the Pt thickness and roughness. The black lines represent simulated curves fitted to the experimental data using the GenX package.}
    \label{Fig.S1}
\end{figure}

\subsection*{Note 2: FIB patterned Nb–Pt–Nb nano Junctions and their superconducting transitions}
\begin{figure}[h!]
    \centering
    \includegraphics[width=1\textwidth]{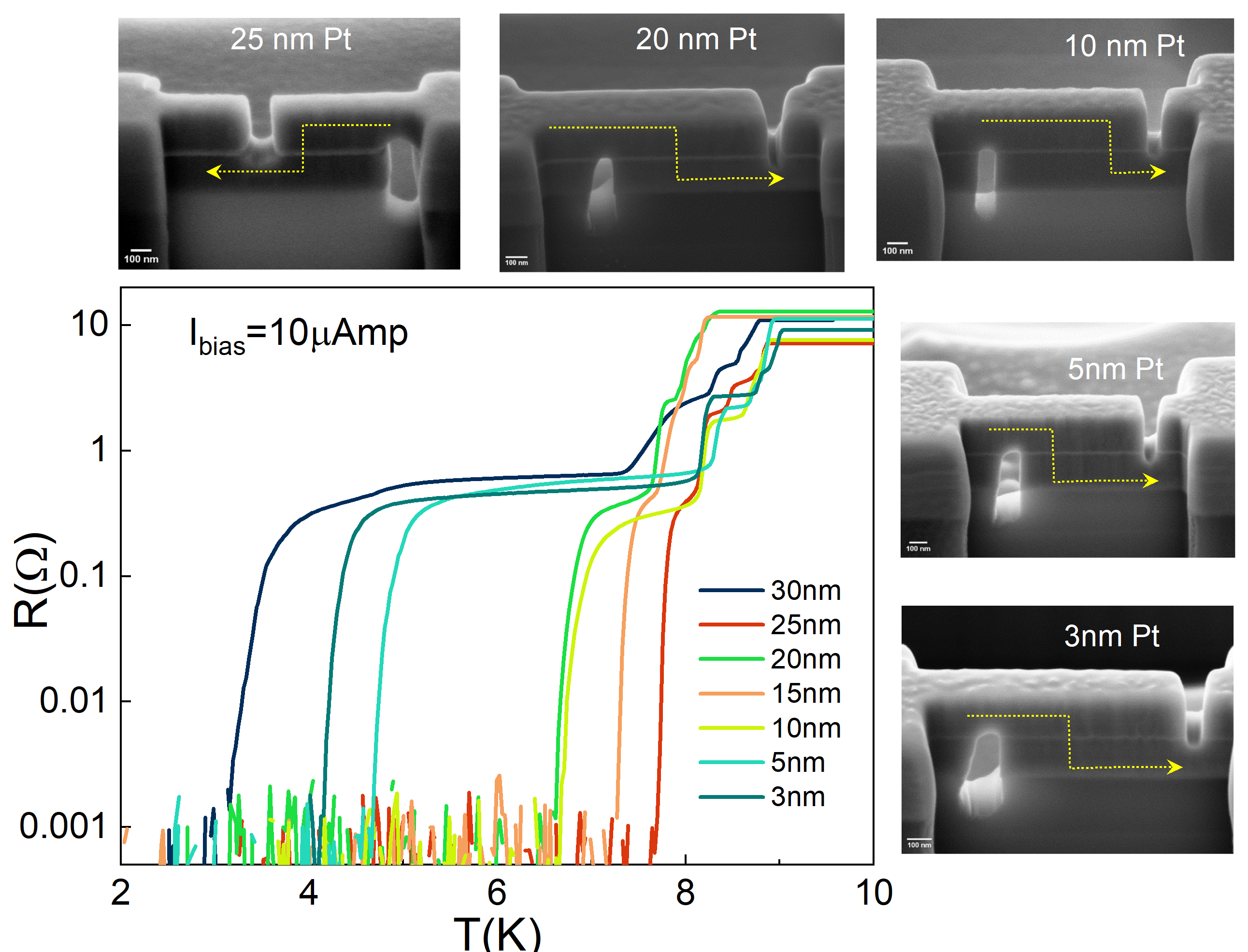} 
    \caption{ Resistance–temperature (R–T) curves for the fabricated Josephson junctions with various Pt thicknesses between 3 nm and 30 nm. The junction transition temperature, show a non-monotonic Pt thickness dependence. All measurements were performed with a fixed bias current of 10 µA, in zero magnetic field condition. Scanning electron micrograph of some representative devices from various thicknesses are also shown. The arrows these images show the current path across the vertical junctions.  }
    \label{Fig.1}
\end{figure}
All junctions were fabricated from trilayers grown under identical growth conditions in the same deposition run. A combination of optical lithography and focused ion beam patterning was used for the fabrication. Especially, the active junction areas were defined by focused Ga ion beam milling of the top and bottom electrodes on either side of the junction area. The edges of the junctions are, therefore, exposed to Ga ion contamination. Depending on the degree of contamination the electrode transition temperature can vary to some extent which may be noticed in the Fig. S2. Some representative electron micrograph images of the fabricated devices with different barrier thickness (between 3nm and 30 nm) are shown in the Fig S2, along with the resistance-temperature ($R$-$T$) characteristics of those devices. A fixed bias current of 10\,$\mu$A was used for all the measurements, in zero magnetic field. Upon cooling, several resistance drops are observed in the the superconducting transitions of the Nb electrodes, corresponding to different widths segments of the electrode necessitated by the focused ion beam fabrication process. The transition at the lowest temperature corresponds to the proximatization of the Josephson junction $T_\text{JJ}$. As the Pt barrier thickness increases, the junction transition temperature $T_\text{JJ}$ does not change monotonically. Instead, $T_\text{JJ}$ increases up to a thickness of approximately 25\,nm and decreases for larger thicknesses, as visible in Fig.S2. For the 15\,nm and 20\,nm samples, a shift in the electrode transition temperature $T_\text{SC}$ was observed. Therefore, the normalized ratio $T_\text{JJ}/T_\text{SC}$ is used in the main text for comparison across different thicknesses.

\subsection*{Note 3: Further data on Josephson diode effect in Nb-Pt-Nb junctions}
\begin{figure}[h!]
    \centering
    \includegraphics[width=1\textwidth]{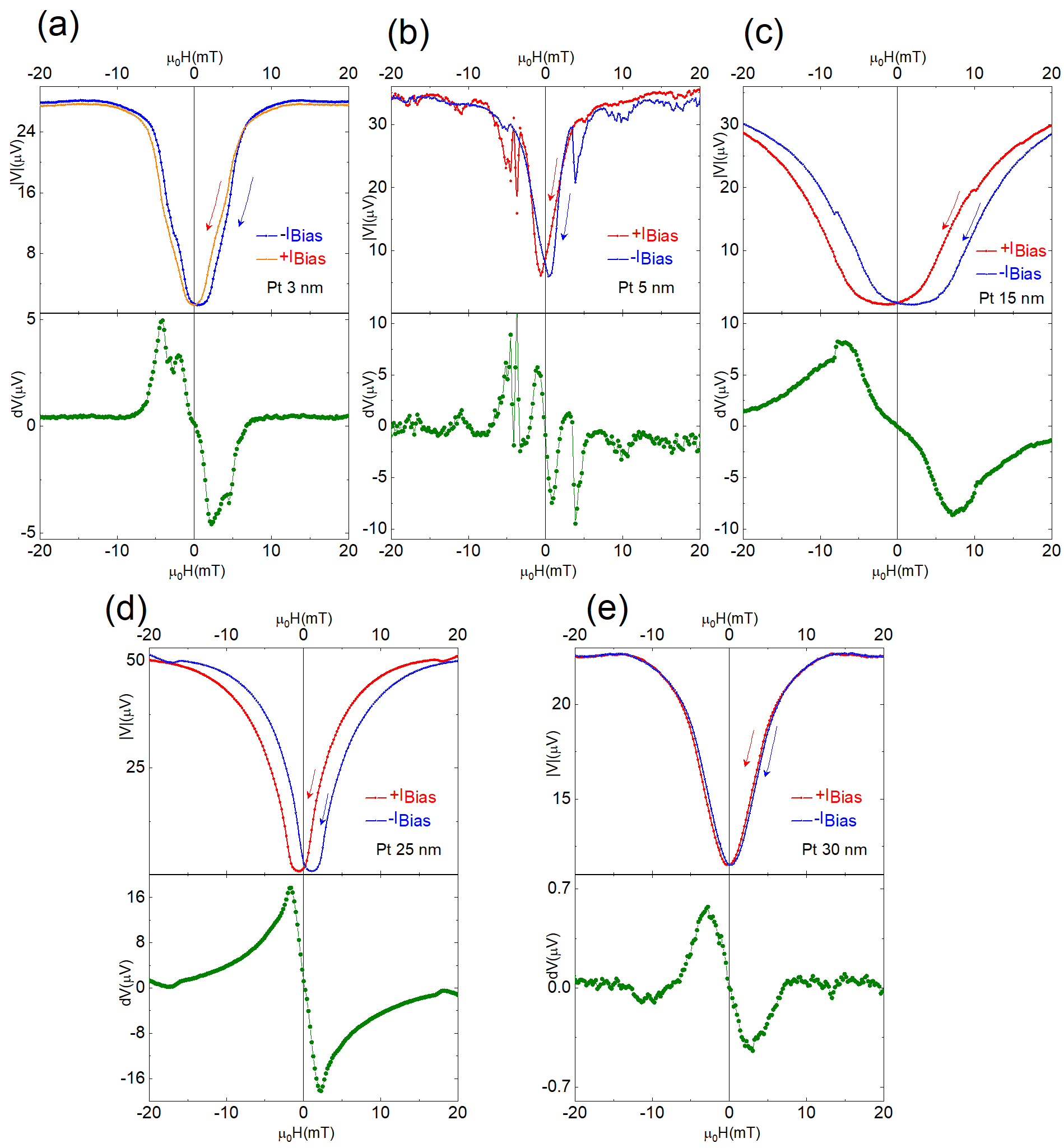} 
    \caption{The field dependence of the voltage across the junctions are plotted for forward and reverse bias currents for Nb-Pt-Nb junctions with various Pt barrier thickness. The corresponding $\Delta$V, calculated from these measurements, are also plotted in each case signifying the field dependence of the diode effect. Signatures of second harmonic Josephson current are apparent in junctions with 3 nm, 5 nm, and 15 nm barriers, as seen in panels a, b, and c, respectively.}
    
\end{figure}
The Josephson diode effect in the series of Nb-Pt-Nb junctions was evaluated through the measurements of field dependent voltage across the junctions. The difference in voltage($\Delta$V) measured with opposite bias currents, was used to evaluate the field dependence of diode effect. Supplementary Fig S3 shows some representative V(H) curves for various thickness of Pt and the corresponding $\Delta$V, measured at a temperature close to 90\% of the junction transition temperature. Clearly the JDE appears to be a consistent feature in junctions across the entire series of Nb-Pt-Nb junctions. The magnitude of the JDE, however varies in a non-monotonic manner as a function of the thickness of the Pt barrier. All V(H) measurements were performed with the same magnetic field sweep direction (positive to negative) to avoid any possible hysteretic artifact, though no hysteresis in V(H) data was observed between forward and reverse sweeps of magnetic field. Typically, Josephson junctions with magnetic barriers show a characteristic hysteresis behavior in the field dependence of critical current, between the forward and reverse field sweeps. This is caused by the underlying magnetic hysteresis of the barrier layer. The Nb-Pt-Nb nano-junctions studied in this work have no magnetic component. Therefore, no hysteresis is expected in the voltage across the junction (V(H)) between forward and reverse field sweeps. A sample V(H) curve, measured with forward and reverse sweeps of magnetic field, is shown in the Fig S4 for a Nb-Pt(30nm)-Nb junction, which shows no discernible hysteresis.   

\begin{figure}[h!]
    \centering
    \includegraphics[width=0.5\textwidth]{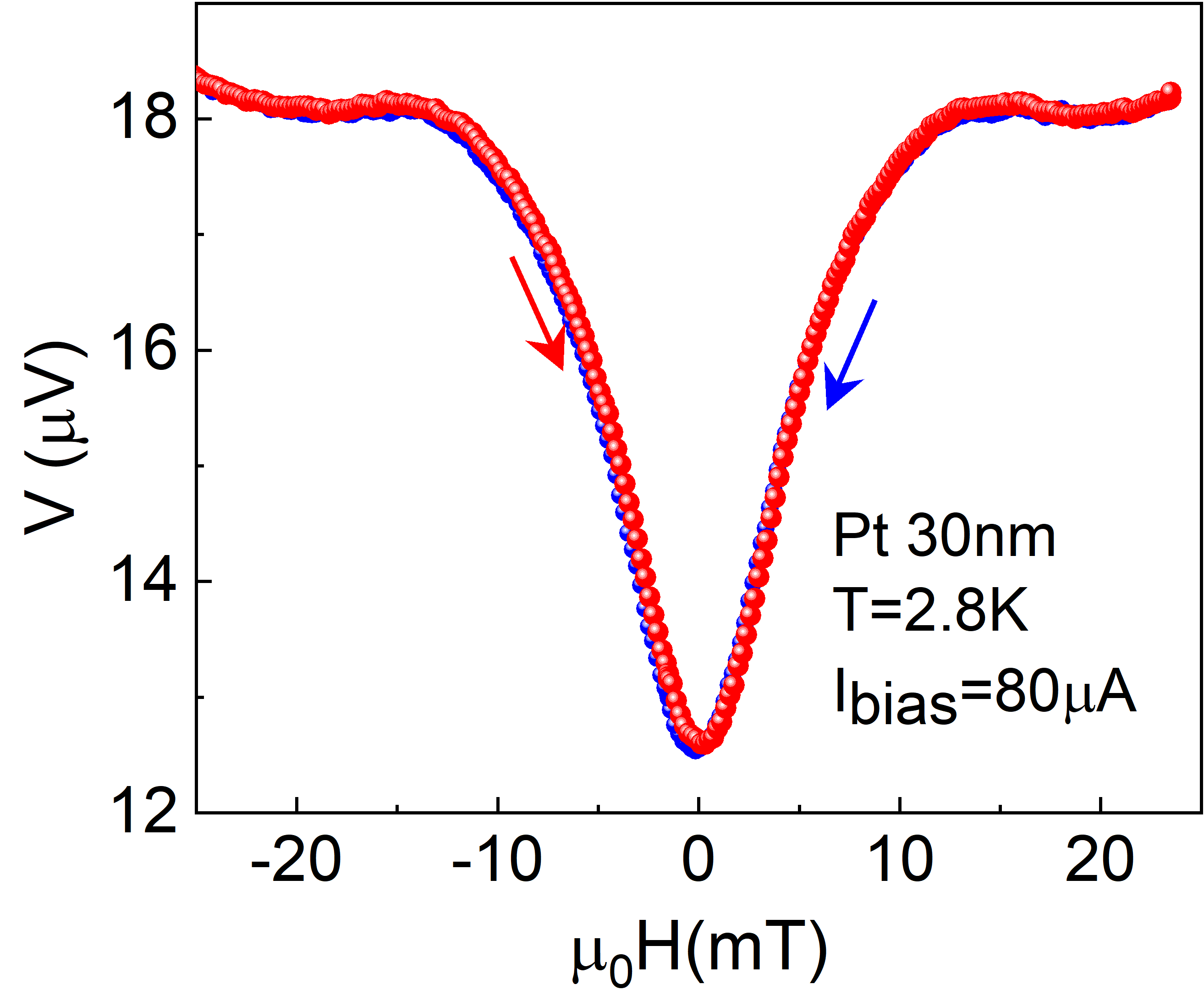} 
    \caption{The field dependence of the voltage across the junctions is plotted for forward and reverse sweep of magnetic field for a Nb-Pt(30nm)-Nb junctions.}
    \label{Fig.S1}
\end{figure}

\subsection*{Note 4: Signature of 0-$\pi$ crossover in the Nb-Pt-Nb junctions}

The other interesting feature of the V(H) data in Fig S3 is the prominent second harmonic features in the case of 3nm and 5nm Pt barrier junctions. As discussed in the main text, the normalized junction transition temperature shows a non-monotonic dependence on the Pt barrier thickness, akin to the case of 0-$\pi$ transitions in magnetic Josephson junctions. A validating feature of the 0-$\pi$ transition point along any parameter axis (such as temperature or barrier thickness )is the appearance of the second harmonic current in the Josephson junction near the 0-$\pi$ transition point. In the main text we have demonstrated the appearance of second harmonic signature in the V(H) curve near the 0-$\pi$ temperature boundary. Along the thickness axis in the Fig 3(b) in the main text, the 0-$\pi$ boundary appears to be close to 5 nm. As a result, the 3 nm and 5 nm junctions show strong second harmonic features in the Fig S3(a) and Fig S3(b), manifested as an additional periodicity near half flux quantum. The 3\,nm Pt junction has a primary Fraunhofer period of approximately 100\,G, while the bump (shown by the arrow) in Fig S3(a) near half flux quantum, corresponds to a second harmonic signature. This junction was measured at 2.9\,K with a bias current of 110\,$\mu$A. Similarly, the 5\,nm Pt junction, has a primary period of $\sim$80 Oe  while the second-harmonic feature is observed near 40 Oe. This data was obtained at 5.6\,K and a bias current of 220\,$\mu$A.
\begin{figure}[h!]
    \centering
    \includegraphics[width=1\textwidth]{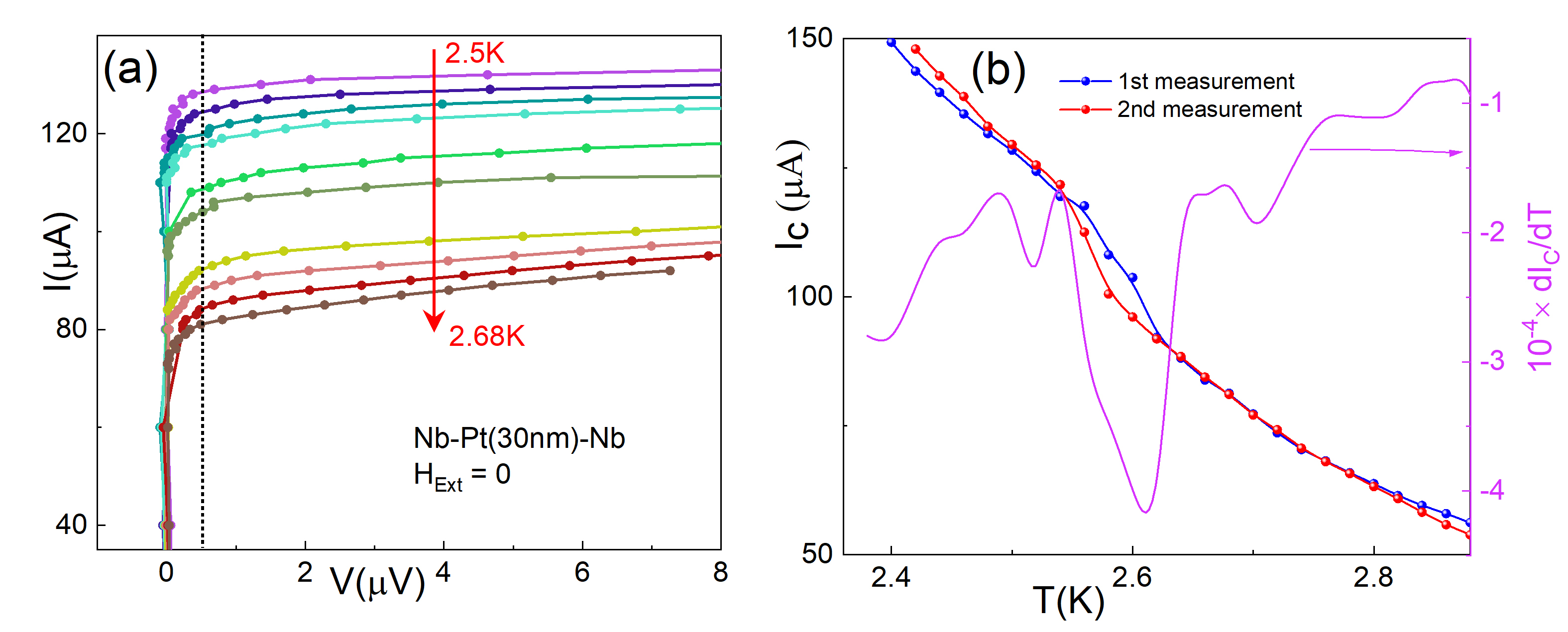} 
   \caption{(a) Current–voltage (IV) characteristics of the 30\,nm Pt junction measured between 2.52\,K and 2.68\,K in steps of 0.02\,K. The dotted vertical line denotes the voltage threshold used to extract the critical current $I_C$. (b) Two independent measurements of temperature dependence of the critical current, $I_c(T)$, extracted from the IV data, are plotted to demonstrate the reproducibility of the drop in the critical current around 2.6 K. The sudden drop in critical current at $\sim$2.6 K is also apparent in the slope of the I$_C$(T) curve plotted on the right hand axis of this figure. }
    \label{}
\end{figure}
In the main text we have also presented the signature of temperature dependent 0-$\pi$ crossover in a Nb-Pt(30nm)-Nb junction as a sharp change in the magnitude of the current I$_C$(T) at 2.6 K. The current-voltage curves across 2.6 K are plotted in the panel (a) of the supplementary Fig S5. The vertical dotted line in this panel is at a voltage of 0.5$\mu$V, which was used as the critical voltage to define the critical current. The panel (b) in Fig S5 shows two independent measurements of the same junction to ascertain the drop in I$_C$ around 2.6 K for this junction and it was found to be consistent in both measurements. The slope of the I$_C$(T) curve has been plotted on the right hand axis of the Fig S5(b), also showing a distinct drop around 2.6 K. The observed second harmonic signature around 2.6 K in the main text Fig 3 (d), corresponding to the drop in the critical current, is a strong indication that the drop corresponds to a 0-$\pi$ crossover point.

The origin of the 0-$\pi$ crossover in the non-magnetic Nb-Pt-Nb junctions is the effective spin moment generated by the triplet component of the superconducting condensate in the Pt barrier layer. The fact that there is no magnetic origin of this effect in our devices can be argued based on the non-hysteretic V(H) curve in Fig S4. We must also mention that we have ruled out the possibility of any magnetic impurity in the barrier which might cause 0-$\pi$ crossover, though the impurity level has to be reasonably high to establish a magnetic exchange. EDX measurements with long acquisition times, however, did not show any signature of magnetic contamination of the junctions as shown in the sample EDX spectrum plotted in the Fig S6. 
\begin{figure}[h]
    \centering
    \includegraphics[width=.4\textwidth]{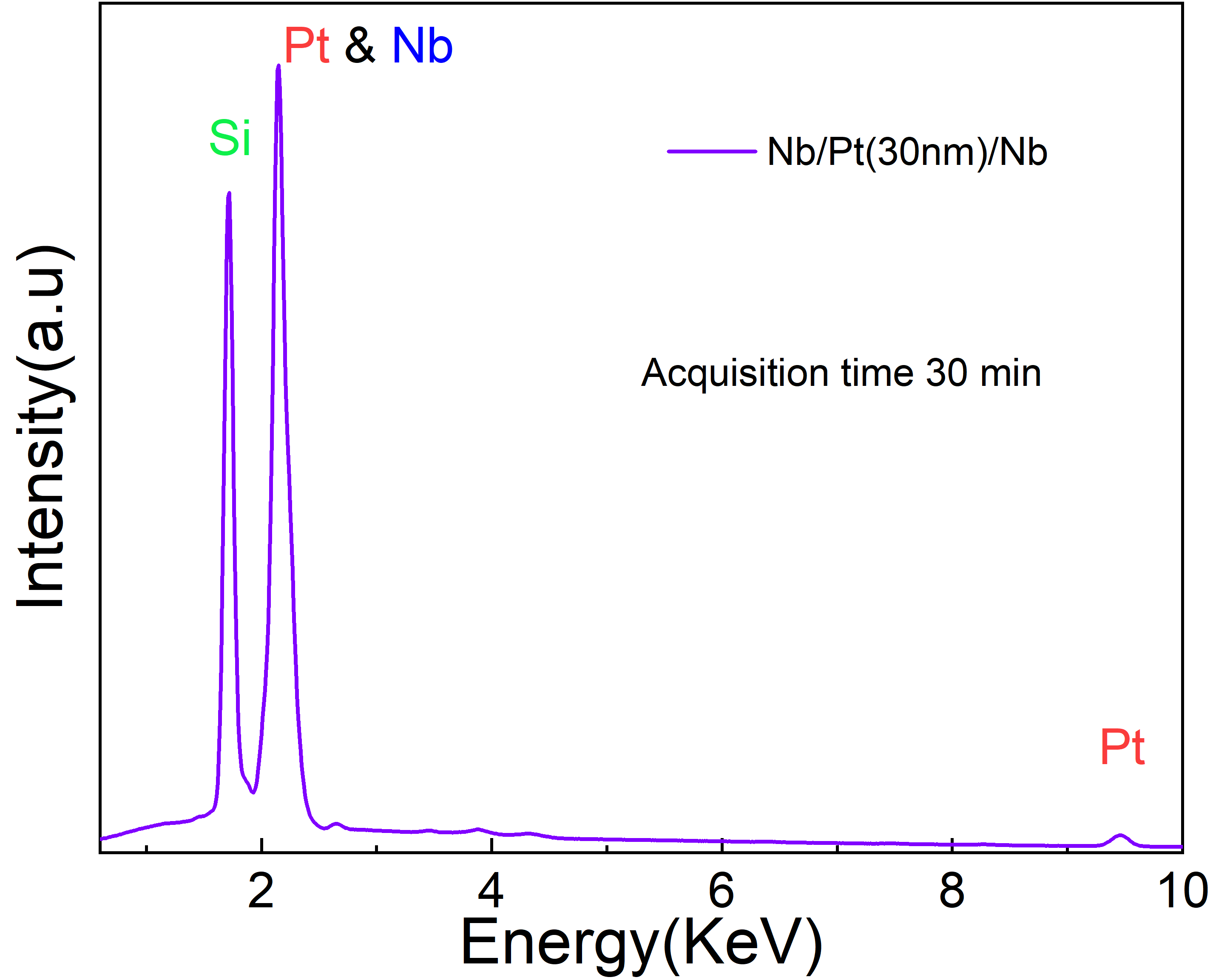} 
   \caption{ Energy dispersive X-ray spectrum of a Nb-Pt-Nb trilayer is shown with a 30 minutes acquisition time, measured upto incident energy of 10 kV.}
    \label{}
\end{figure}

\newpage
\subsection*{Note 5: Further data on spin valve type response in Nb-Pt-Ni-Nb junction}
In the main text we have discussed about the spin-valve like response of a Nb-Pt-Ni-Nb junction, above the full proximatization temperature of the (Ni-Pt) bilayer barrier. The triplet induced spin moment in Pt layer and the magnetization of the Ni layer form a spin-valve like arrangement. The magnitude of the magnetoresistance peaks for the forward and reverse field sweeps were found to be slightly different. The difference ($\Delta$R) between the forward and reverse MR exhibits opposite trends upon reversal of the bias current direction, as shown for 4.8 K in the Fig 4(c) in the main text. In the supplementary Fig S7 we plot the $\Delta$R measured with the same bias currents at 5.4 K and 6.5 K. Though very small in magnitude, in all cases we observe the current dependent reversal in the magnitude of $\Delta$R. This observation confirms the existence of a spin moment in the Pt layer in proximity with the superconducting Nb layer.

\begin{figure}[h!]
    \centering
    \includegraphics[width=1\textwidth]{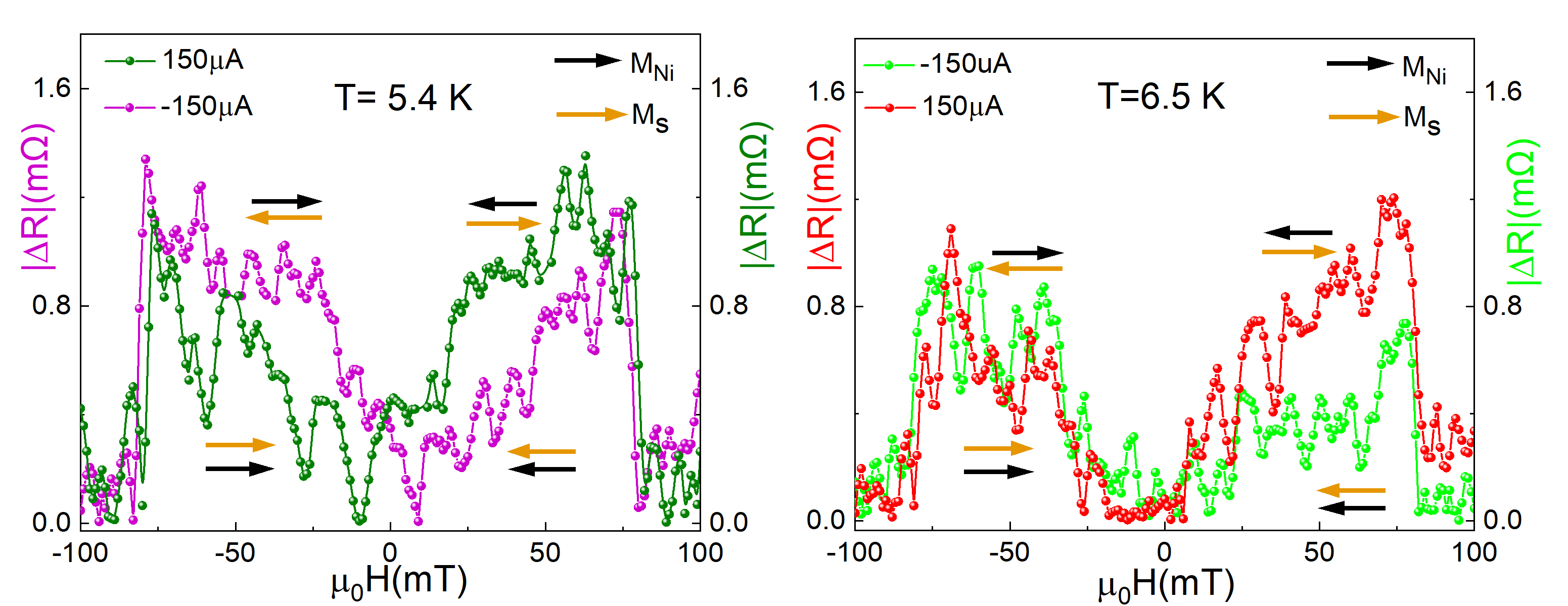} 
   \caption{The difference between the resistance curves for forward and reverse field sweeps are plotted as a function of magnetic field at temperatures of 5.4 K and 6.5 K.   }
    \label{}
\end{figure}

\newpage

\section*{Note 6: Electronic Structure calculation of the Nb/Pt Heterostructure}
The first-principles~\cite{Hohenberg1964,Kohn1965} electronic structure calculations are discussed in the methods section of the main paper.
Constructing a coherent Nb/Pt interface is nontrivial because of the substantial lattice mismatch and crystallographic incompatibility between bcc Nb(110) and fcc Pt(111) surfaces. Although the experimentally investigated trilayers are polycrystalline rather than epitaxial, these interface orientations were selected based on the preferential growth directions of the sputtered Nb and Pt thin films, in our case. Bulk Nb crystallizes in a body-centered cubic structure with an 
optimized lattice constant of $a_{\mathrm{Nb}} = 3.32$~\text{\AA}) ~\cite{nb,Ness2022}, whereas Pt adopts a face-centered cubic structure with an optimized lattice constant 
of $a_{\mathrm{Pt}} = 3.91$~\text{\AA})~\cite{Barreteau_2012,PhysRevB.70.235423}. 
The corresponding optimized bulk crystal structures of bulk Nb and Pt are shown in Supplementary Fig. S8(a,b).\\

To construct a commensurate interfacial heterostructure, the primitive surface cells were expanded into $\sqrt{3}\times\sqrt{3}\times1$ and $\sqrt{2}\times\sqrt{2}\times1$ supercells for Nb(110) and Pt(111), respectively. This supercell matching reduces the residual in-plane lattice mismatch to approximately $3.48$\%, enabling construction of a structurally coherent interface. The Nb and Pt slabs were subsequently combined to 
form the heterostructure shown schematically in Fig. 4(d) of the main text, where effective slab thicknesses are approximately 4.7~\AA\ for Nb and 4.4~\AA\ for Pt, respectively. A vacuum spacing exceeding $20~\text{\AA}$ was introduced along the surface-normal direction to eliminate spurious interactions between periodic images. The equilibrium interfacial spacing between Nb and Pt layers was determined by minimizing the interface formation energy,
\begin{equation}
F.E.=E_{\mathrm{Nb/Pt}}-E_{\mathrm{Nb}}-E_{\mathrm{Pt}}~,
\end{equation}
where $E_{\mathrm{Nb/Pt}}$ is the total energy of the combined heterostructure, 
while $E_{\mathrm{Nb}}$ and $E_{\mathrm{Pt}}$ correspond to the formation energies of the isolated Nb and Pt slabs, evaluated using identical in-plane lattice parameters and atomic configurations. As shown in Supplementary Fig. S8(c), the formation energy exhibits a pronounced minimum at an interfacial separation of approximately $3.2~\text{\AA}$, identifying the energetically preferred interface geometry. 
The negative formation energy further confirms that the Nb(110)/Pt(111) interface is thermodynamically stable and electronically favorable.

\begin{figure}
\centering
\includegraphics[width=1\linewidth]{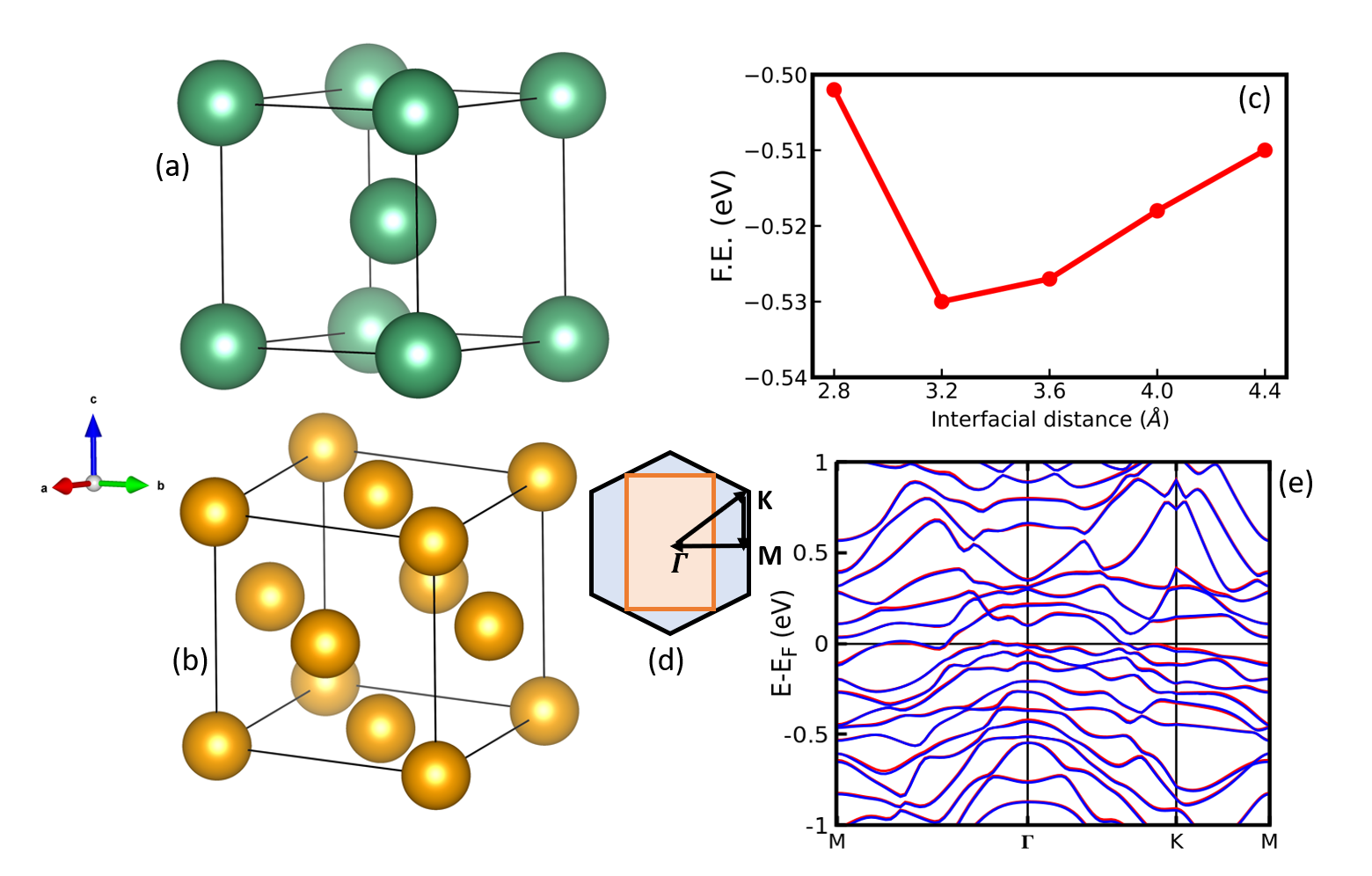}
\caption{(a) Optimized bulk crystal structure of bcc Nb. (b) Optimized bulk 
crystal structure of fcc Pt. (c) Formation energy ((F.E.)) of the Nb(110)/Pt(111) heterostructure as a function of the interfacial separation between Nb and Pt layers, exhibiting a minimum near the equilibrium interface distance. (d) Two-dimensional hexagonal Brillouin zone indicating the high-symmetry path $M \rightarrow \Gamma \rightarrow K \rightarrow M$ used for the electronic band-structure calculations. (e) Spin-polarized electronic band structure is calculated without SOC along the indicated $k$-path. The red and blue curves denote the two spin-degenerate channels, while the dashed horizontal line represents the Fermi level $E_{\mathrm{F}} = 0$ eV).}

\end{figure}

Spin-resolved band projections and momentum-dependent spin expectation values further reveal interfacial spin-polarization channels arising from the combined effects of strong Pt spin–orbit interaction and broken inversion symmetry. 
Electronic dispersions were evaluated along the high-symmetry directions of the two-dimensional surface Brillouin zone shown in Supplementary Fig. S8(d). To elucidate the role of the Pt barrier in generating nonreciprocal superconducting transport, electronic band structures of the Nb/Pt heterostructure were computed both with and without spin–orbit coupling (SOC). In the absence of SOC, the heterostructure remains nonmagnetic and the electronic bands are spin degenerate, reflecting preservation 
of spin-rotational symmetry, as shown in Supplementary Fig. S8(e). Upon inclusion of SOC within the non-collinear formalism using the second-variational approach, inversion-symmetry breaking at the Nb/Pt interface lifts the spin degeneracy of bands near the Fermi level and generates momentum-dependent spin textures and interfacial spin polarization arising 
from the strong atomic SOC of Pt. To accurately capture these interfacial SOC effects, crystalline symmetry was disabled during the SOC calculations.

\newpage
\putbib[supp]   

\end{bibunit}

\end{document}